\documentclass{article}

\PassOptionsToPackage{numbers,sort&compress}{natbib}
\usepackage[preprint]{neurips_2023}

\usepackage{cite}
\usepackage[utf8]{inputenc} 
\usepackage[T1]{fontenc}    
\usepackage[colorlinks=false]{hyperref}
\hypersetup{
  hidelinks,
  citecolor=black,
  linkcolor=black,
  urlcolor=black}
\usepackage{booktabs}       
\usepackage{amsfonts}       
\usepackage{nicefrac}       
\usepackage{microtype}      
\usepackage{graphicx}
\usepackage{enumitem}

\NewDocumentCommand{\rot}{O{45} O{1em} m}{\makebox[#2][l]{\rotatebox{#1}{#3}}}%

\usepackage[table]{xcolor}         

\usepackage{arydshln}

\usepackage[acronym,shortcuts]{glossaries}
\newacronym{CNN}{CNN}{convolutional neural network}
\newacronym{ANN}{ANN}{artificial neural network}
\newacronym{AI}{AI}{artifical intelligence}
\newacronym{DoG}{DoG}{Difference of Gaussian}
\newacronym{RF}{RF}{receptive field}
\newacronym{V1}{V1}{primary visual cortex}

\title{Explaining V1 Properties with a Biologically Constrained Deep Learning Architecture}

\author{%
  Galen Pogoncheff \\
  Department of Computer Science\\
  University of California, Santa Barbara\\
  Santa Barbara, CA 93106 \\
  \texttt{galenpogoncheff@ucsb.edu} \\
  \And
  Jacob Granley \\
  Department of Computer Science\\
  University of California, Santa Barbara\\
  Santa Barbara, CA 93106 \\
  \texttt{jgranley@ucsb.edu} \\
  \AND
  Michael Beyeler \\
  Department of Computer Science, \\
  Department of Psychological and Brain Sciences \\
  University of California, Santa Barbara\\
  Santa Barbara, CA 93106 \\
  \texttt{mbeyeler@ucsb.edu} \\
}

\begin{document}

\maketitle

\begin{abstract}
  
  \Acfp{CNN} have recently emerged as promising models of the ventral visual stream, despite their lack of biological specificity.
  While current state-of-the-art models of the \acf{V1} have surfaced from training with adversarial examples and extensively augmented data, these models are still unable to explain key neural properties observed in V1 that arise from biological circuitry.
  To address this gap, we systematically incorporated neuroscience-derived architectural components into CNNs to identify a set of mechanisms and architectures that comprehensively explain neural activity in V1.
  We show drastic improvements in model-V1 alignment driven by the integration of architectural components that simulate center-surround antagonism, local receptive fields, tuned normalization, and cortical magnification.  
  Upon enhancing task-driven CNNs with a collection of these specialized components, we uncover models with latent representations that yield state-of-the-art explanation of V1 neural activity and tuning properties.
  Our results highlight an important advancement in the field of NeuroAI, as we systematically establish a set of architectural components that contribute to unprecedented explanation of V1.
  The neuroscience insights that could be gleaned from increasingly accurate in-silico models of the brain have the potential to greatly advance the fields of both neuroscience and \acf{AI}.
\end{abstract}

\section{Introduction}

Advances in neuroscience have long been proposed as essential to realizing the next generation of \ac{AI}.
Many influential deep learning architectures and mechanisms that are widely used today (e.g, convolutional neural networks and mechanisms of attention) owe their origins to biological intelligence.
Despite decades of research into computational models of the visual system, our understanding of its complexities remains far from complete.
Existing neuroscientific models of the visual system are often founded upon empirical observations from relatively small datasets, and are therefore unlikely to capture the true complexity of the visual system.
While these models have successfully explained many properties of neural response to simple stimuli, their simplicity does not generalize to complex image stimuli \citep{carandini_we_2005}.

Following their astounding success in computer vision, task-driven \acp{CNN} have recently been proposed as candidate models of the ventral stream in primate visual cortex \citep{yamins_hierarchical_2013,yamins_performance-optimized_2014,yamins_using_2016,majaj_simple_2015,cadena_deep_2019}, offering a path towards models that can explain hidden complexities of the visual system and generalize to complex visual stimuli.
Through typical task-driven training alone, representations that resemble neural activity at multiple levels of the visual hierarchy have been observed in these models.
With the emergence of such properties, \acp{CNN} are already being used to enhance our knowledge of processing in the ventral stream \citep{bashivan_neural_2019}.

Despite these advancements, \acp{CNN} that achieve state-of-the-art brain alignment are still unable to explain many properties of the visual stream.
Most traditional \acp{CNN} omit many well known architectural and processing hallmarks of the primate ventral stream that are likely key to the development of artificial neural nets that help us decipher the neural code.
The development of these mechanisms remains an open challenge.
A comprehensive understanding of the visual stream could in turn contribute to significant leaps in \Acs{AI} -- a long-established goal of NeuroAI research.

In this work, we take a systematic approach to analyzing the hallmarks of the primate ventral stream that improve model-brain similarity of \acp{CNN}.
We formulate architectural components that simulate these processing hallmarks within \acp{CNN} and analyze the population and neuron level response properties of these networks, as compared to empirical data recorded in primates.
In specific:
\begin{itemize}[topsep=0pt,leftmargin=15pt,parsep=0pt]
    \item We introduce architectural components based on neuroscience foundations that simulate
    cortical magnification, center-surround antagonism, local filtering, and tuned divisive normalization.
    \item We systematically analyze how these architectural components lead to latent representations that better explain primate V1 activitiy with multifaceted Brain-Score analyses. We identify center-surround antagonism, local filtering, tuned normalization, and cortical magnification as complementary ways to improve V1 alignment.
    \item We enrich the classic ResNet50 architecture with these architectural components and show that the resulting network achieves top V1-overall score on the integrative Brain-Score benchmark. An ablation study reveals the importance of each component and the benefits of combining multiple of these components into a single neuro-constrained \acs{CNN}.
\end{itemize}

\section{Background and Related Work}

\paragraph{Model-Brain Alignment}

One central challenge in the field of NeuroAI is the development of computational models that can effectively explain the neural code.
To achieve this goal, artificial neural networks must be capable of accurately predicting the behavior of individual neurons and neural populations in the brain.
The \acf{V1} is one of the most well studies areas of the visual stream, with modeling efforts dating back to at least 1962 \citep{hubel_receptive_1962}---yet many deep learning models still fall short in explaining its neural activity.

The Brain-Score integrative benchmark \citep{schrimpf_brain-score_2020} has recently emerged as a valuable tool for assessing the capabilities of deep learning models to explain neural activity in the visual system.
This suite of benchmarks integrates neural recording and behavioral data from a collection of previous studies and provides standardized metrics for evaluating model explainability of visual areas V1, V2, V4, and IT, as well as additional behavioral and engineering benchmarks. 

Although \acp{CNN} draw high-level inspiration from neuroscience, current architectures (e.g., ResNet \citep{he_deep_2016} and EfficientNet \citep{tan_efficientnet_2019}) bear little resemblance to neural circuits in the visual system.
While such differences may not necessarily hinder object recognition performance, these networks still fall short in mimicking many properties of highly capable visual systems.
Although there may be many paths towards next-generation \acs{AI}, foundational studies that have successfully merged foundations of neuroscience and \acs{AI} have shown promising improvements to traditional \acp{ANN} \citep{li_learning_2019,dapello_simulating_2020,reddy_biologically_2020}.

\paragraph{Center-Surround Antagonism}
As early as in the retina, lateral inhibitory connections establish a center-surround antagonism in the \ac{RF} of many retinal cell types, which is preserved by neurons in the lateral geniculate nucleus and the visual cortex.
In the primate visual stream, this center-surround antagonism is thought to facilitate edge detection, figure-ground segregation, depth perception, and cue-invariant object perception \citep{allman_stimulus_1985,knierim_neuronal_1992,walker_asymmetric_1999,shen_cue-invariant_2007}, and is therefore a fundamental property of visual processing.

\begin{figure}[tb!]
    \centering
    \includegraphics[width=\linewidth]{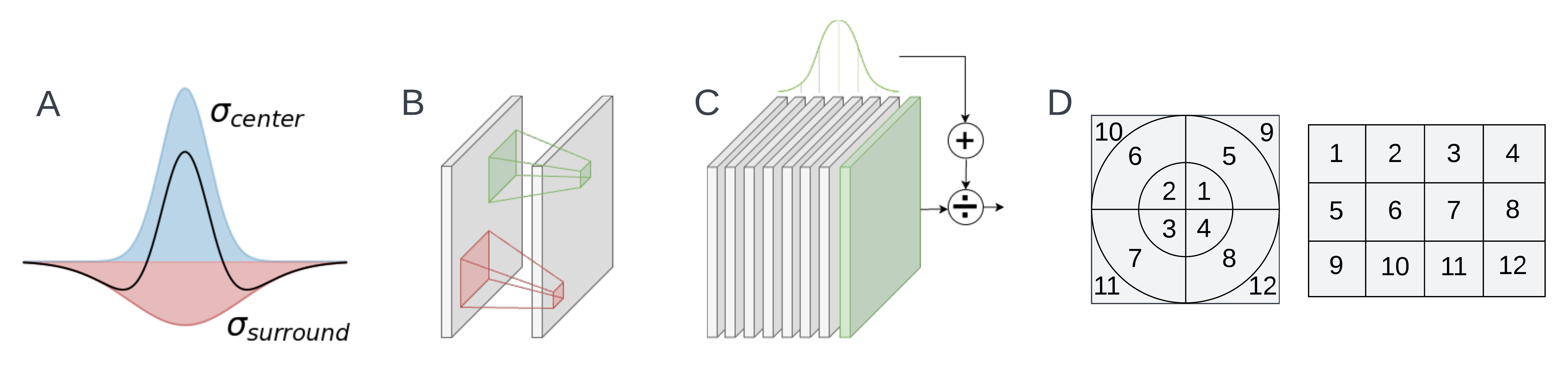}
    \caption{Design patterns of neuro-constrained architectural components.
    A) Difference of Gaussian implements a center-surround receptive field.
    B) Local receptive fields of two neurons without weight sharing.
    C) Tuned divisive normalization inhibits each feature map by a Gaussian-weighted average of competing features.
    D) Log-polar transform simulating cortical magnification}
    \label{fig:bg-components}
\end{figure}

Center-surround \acp{RF} are a common component of classical neuroscience models \citep{deangelis_length_1994,sceniak_contrasts_1999,sceniak_visual_2001}, where they are typically implemented using a \ac{DoG} that produces an excitatory peak at the \ac{RF} center with an inhibitory surround (Fig.~\ref{fig:bg-components}A).
Although deep \acp{CNN} have the capacity to learn center-surround antagonism, supplementing traditional convolutional kernels with fixed-weight \ac{DoG} kernels has been demonstrated to improve object recognition in the context of varied lighting, occlusion, and noise \citep{hasani_surround_2019, babaiee_-off_2021}.

\paragraph{Local Receptive Fields}
The composition of convolutional operations in \acp{CNN} enables hierarchical processing and translation equivariance, both of which are fundamental to core object recognition in the primate ventral visual stream.
However, the underlying mechanism through which this is achieved is biologically implausible, as kernel weights are shared among downstream neurons.
Though locally connected neural network layers can theoretically learn the same operation, traditional convolutions are typically favored in practice for their computational efficiency and performance benefits.
However, local connectivity is a ubiquitous pattern in the ventral stream (Fig.~\ref{fig:bg-components}B), and visual processing phenomena (e.g., orientation preference maps \citep{koulakov_orientation_2001}) have been attributed to this circuitry pattern.
In artificial neural systems, Lee \emph{et al.} \citep{lee_topographic_2020} observed the emergence of topographic hallmarks in the inferior temporal cortex when encouraging local connectivity in \acp{CNN}.
Pogodin \emph{et al.} \citep{pogodin_towards_2021} considered the biological implausibility of \acp{CNN} and demonstrated a neuro-inspired approach to reducing the performance gap between traditional \acp{CNN} and locally-connected networks, meanwhile achieving better alignment with neural activity in primates.

\paragraph{Divisive Normalization}

Divisive normalization is wide-spread across neural systems and species \citep{carandini_normalization_2012}. 
In early visual cortex, it is theorized to give rise to well-documented physiological phenomena, such as response saturation, sublinear summation of stimulus responses, and cross-orientation suppression \citep{heeger_recurrent_2020}.

In 2021, Burg and colleagues \citep{burg_learning_2021} introduced an image-computable divisive normalization model in which each artificial neuron was normalized by weighted responses of neurons with the same receptive field.
In comparison to a simple 3-layer \acrshort{CNN} trained to predict the same stimulus responses, their analyses revealed that cross-orientation suppression was more prevalent in the divisive normalization model than in the \acrshort{CNN}, suggesting that divisive normalization may not be inherently learned by task-driven \acp{CNN}.
In a separate study, Ciricione \emph{et al.} \citep{cirincione_implementing_nodate} showed that simulating divisive normalization within a \acs{CNN} can improve object recognition robustness to image corruptions and enhance alignment with certain tuning properties of primate \acs{V1}.

\paragraph{Tuned Normalization/Cross-Channel Inhibition}
While it is not entirely clear whether divisive normalization should be performed across space and/or across channels (implementations vary widely), Rust \emph{et al.} \citep{rust_how_2006} demonstrated that many response properties of motion-selective cells in the middle temporal area, such as motion-opponent suppression and response normalization, emerge from a mechanism they termed ``tuned normalization''.
In this scheme, a given neuron is normalized by a pool of neurons that share the same receptive field but occupy a different region in feature space.
We adopt this idea in the present work (Fig.~\ref{fig:bg-components}C), hypothesizing that enforcing feature-specific weights in the pooling signal might enable a deep net to learn ``opponent suppression'' signals, much like cross-orientation signals found in biological V1 \citep{morrone_functional_1982,deangelis_organization_1992}.

\paragraph{Cortical Magnification}
In many sensory systems, a disproportionately large area of the cortex is dedicated to processing the most important information.
This phenomenon, known as cortical magnification, reflects the degree to which the brain dedicates resources to processing sensory information accompanying a specific sense.
In the primary visual cortex, a larger proportion of cortical area processes visual stimuli presented at the center of the visual field as compared to stimuli at greater spatial eccentricities \citep{daniel_representation_1961}.
The relationship between locations in the visual field and corresponding processing regions in the visual cortex has commonly been modeled with a log-polar mapping (Fig.~\ref{fig:bg-components}D) or derivations thereof \citep{schwartz_spatial_1977,schwartz_computational_1980,schwartz_computational_1994,polimeni_multi-area_2006}.

Layers of artificial neurons of traditional \acp{CNN} have uniform receptive field sizes and do not exhibit any sort of cortical magnification, failing to capture these distinctive properties of neuronal organization in the primary visual cortex.
Recent works have demonstrated that introducing log polar-space sampling into \acp{CNN} can give rise to improved invariance and equivariance to spatial transformations \citep{carlos_esteves_polar_2018, henriques_warped_2021} and adversarial robustness \citep{kiritani_recurrent_2020}.

\section{Methods}

\begin{figure}[t!]
    \centering
    \includegraphics[width=\linewidth]{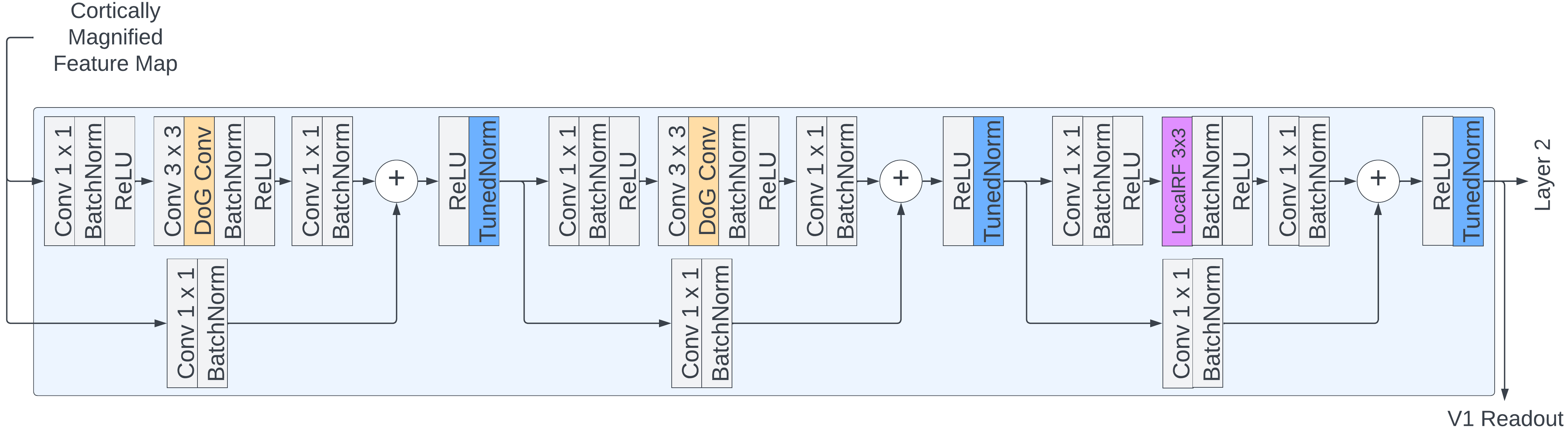}
    \caption{ResNet50 layer 1, supplemented with neuro-constrained architectural components.  Throughout the the modified layer 1, \acf{V1} activity is modeled with cortical magnification, center-surround convolutions, tuned normalization, and local receptive field layers.  Layer 1 output units are treated as artificial \acs{V1} neurons.}
    \label{fig:architecture}
\end{figure}

\subsection{Neuro-Constrained CNN Architecture}

Given previous state-of-the-art V1 alignment scores achieved with ResNet50 \citep{dapello_simulating_2020}, we adopted this architecture as our baseline and test platform.
However, the architectural components that we considered in the work are modular and can be integrated into general \acp{CNN} architectures.
The remainder of this subsection details each architectural component and its integration into a neuro-constrained ResNet.
In all experiments, we treated the output units from ResNet50 layer 1 as ``artificial \acs{V1}'' neurons (refer to Section~\ref{subsec:training-evaluation} for layer selection criteria).
Fig.~\ref{fig:architecture} depicts ResNet50 layer 1 after enhancement with neuroscience-based architectural components.

\paragraph{Center-Surround Antagonism}

Center-surround \acs{ANN} layers are composed  of \acrshort{DoG} kernels of shape $(c_i \times c_o \times k \times k)$, where $c_i$ and $c_o$ denote the number of input and output channels, respectively, and $k$ reflects the height and width of each kernel.
These \acs{DoG} kernels (Fig.~\ref{fig:bg-components}A) are convolved with the pre-activation output of a standard convolution.
Each DoG kernel, $DoG_i$ is of the form
\begin{equation}
    \mathrm{DoG}_i(x,y) =   \frac{\alpha}{2 \pi \sigma_{i,\mathrm{center}}^2} \exp \Big( -\frac{x^2 + y^2}{2 \sigma_{i,\mathrm{center}}^2} \Big) - \frac{\alpha}{2 \pi \sigma_{i,\mathrm{surround}}^2} \exp \Big( -\frac{x^2 + y^2}{2 \sigma_{i,\mathrm{surround}}^2} \Big),
\end{equation}
where $\sigma_{i,\mathrm{center}}$ and $\sigma_{i,\mathrm{surround}}$ were the Gaussian widths of the center and surround, respectively ($\sigma_{i,\mathrm{center}} < \sigma_{i,\mathrm{surround}}$), $\alpha$ was a scaling factor, and $(x,y) := (0,0)$ at the kernel center.
For $\alpha > 0$, the kernel will have an excitatory center and inhibitory surround while $\alpha < 0$, results in a kernel with inhibitory center and excitatory surround.
Novel to this implementation, each \acs{DoG} kernel has learnable parameters, better accommodating the diverse tuning properties of neurons within the network.
As in \citep{hasani_surround_2019, babaiee_-off_2021}, these \acrshort{DoG} convolutions were only applied to a fraction of the input feature map.
Specifically, we applied this center-surround convolution to one quarter of all $3 \times 3$ convolutions in  layer 1 of our neuro-constrained ResNet50.

\paragraph{Local Receptive Fields}
In an effort to untangle the effects of local connectivity on brain alignment, we modified the artificial \acs{V1} layer by substituting the final $3 \times 3$ convolution of ResNet50 layer 1 with a $3 \times 3$ locally connected layer in isolation.
This substitution assigns each downstream neuron its own filter while preserving its connection to upstream neurons (Fig.~\ref{fig:bg-components}B), following the pattern in \citep{pogodin_towards_2021}.

\paragraph{Divisive Normalization}
We consider the divisive normalization block proposed in \citep{cirincione_implementing_nodate} which performs normalization both spatially and across feature maps using learned normalization pools.
Following our experimental design principle of selectively modifying the network in the vicinity of the artificial \acs{V1} neurons, we added this divisive normalization block after the non-linear activation of each residual block in ResNet50 layer 1.

\paragraph{Tuned Normalization}
We devised a novel implementation of tuned normalization inspired by models of opponent suppression \citep{morrone_functional_1982,deangelis_length_1994,rust_how_2006}.
In this scheme, a given neuron is normalized by a pool of neurons that share the same receptive field but occupy a different region in feature space (Fig.~\ref{fig:bg-components}C), as in \citep{burg_learning_2021,cirincione_implementing_nodate}.
Unlike the learned, weighted normalization proposed in \citep{burg_learning_2021}, tuned inhibition was encouraged in our implementation by enforcing that each neuron was maximally suppressed by a neuron in a different region of feature space, and that no other neuron is maximally inhibited by activity in this feature space.
Letting $x^{c}_{i,j}$ denote the activity of the neuron at spatial location $(i,j)$ and channel $c \in [1,C]$ after application of a non-linear activation function.
The post-divisive normalization state of this neuron, $x'^{c}_{i,j}$, is given by:
\begin{equation}
    x'^{c}_{i,j} = \frac{x^{c}_{i,j}}{1 + \sum_{k} p_k x^{c_k}_{i,j}},
\end{equation}
where $p_{c,1}, \ldots, p_{c,C}$ defines a Gaussian distribution with variance $\sigma^2_{c}$ centered at channel $(c + \frac{C}{2}) \text{ mod } C$.
By defining $\sigma^2_{c}$ as a trainable parameter, task-driven training would optimize whether each neuron should be normalized acutely or broadly across the feature space.

As this mechanism preserves the dimension of the input feature map, it can follow any non-linear activation function of the core network without further modification to the architecture.
Similar to the divisive normalization block, tuned normalization was added after the non-linear activation of each residual block in ResNet50 layer 1 in our experiments.

\paragraph{Cortical Magnification}

Cortical magnification and non-uniform receptive field sampling was simulated in \acp{CNN} using a differentiable polar sampling module (Fig.~\ref{fig:bg-components}D).
In this module, the spatial dimension of an input feature map are divided into polar regions defined by discrete radial and angular divisions of polar space.
In particular, we defined a discrete polar coordinate system partitioned in the first dimension by radial partitions $r_0, r_1, ..., r_m$ and along the second dimension by angular partitions $\theta_0, \theta_1, ..., \theta_n$.
Pixels of the input feature map that are located within the same polar region (i.e., are within the same radial bounds $[r_i, r_{i+1})$ and angular bounds $[\theta_j, \theta_{j+1})$) are pooled and mapped to coordinate $(i,j)$ of the original pixel space (Fig.~\ref{fig:bg-components}D) \citep{blackburn_simple_nodate}.
Pixels in the output feature map with no associated polar region were replaced with interpolated pixel values from the same radial bin.
By defining the spacing between each concentric radial bin to be monotonically increasing (i.e., for all $i \in [1, m-1]$, $(r_i - r_{i-1}) \leq (r_{i+1} - r_{i})$), visual information at lower spatial eccentricities with respect to the center of the input feature map consumes a larger proportion of the transformed feature map than information at greater eccentricities.

A notable result of this transformation is that any standard 2D convolution, with a kernel of size $k \times k$, that is applied to the the transformed features space is equivalent to performing a convolution in which the kernel covers a $k \times k$ contiguous region of polar space and strides along the angular and radial axes.
Reflective padding was used after this transformation to enable a periodic stride along the angular axis at each radial position \citep{carlos_esteves_polar_2018}.
Furthermore, downstream artificial neurons which process information at greater spatial eccentricities obtain larger receptive fields.
Treating the \acs{CNN} as a model of the ventral visual stream, this polar transformation immediately preceded ResNet50 layer 1, where \acs{V1} representations were assumed to be learned.

\subsection{Training and Evaluation}
\label{subsec:training-evaluation}

\paragraph{Training Procedure}
V1 alignment was evaluated for ImageNet-trained models \citep{deng_imagenet_2009}.
Training and validation images were downsampled to a resolution of $64 \times 64$.
Each model of this evaluation was randomly initialized and trained for 100 epochs with an initial learning rate of $0.1$ (reduced by a factor of $10$ at epochs $60$ and $80$, where validation set performance was typically observed to plateau), and a batch size of $128$.

We additionally benchmarked each neuro-constrained model on the Tiny-ImageNet-C dataset to study the effect of V1 alignment on object recognition robustness \citep{hendrycks_benchmarking_2019} (evaluation details provided in Appendix~\ref{app:tinycorruptions}).
Tiny-ImageNet-C was used as an alternate to ImageNet-C given that the models trained here expected $64 \times 64$ input images and downsampling the corrupted images of ImageNet-C would have biased our evaluations. 
ImageNet pre-trained models were fine-tuned on Tiny-ImageNet prior to this evaluation.
As a given model will learn alternate representations when trained on different datasets (thereby resulting in V1 alignment differences), we methodologically froze all parameters of each ImageNet trained model, with the exception of the classification head, prior to 40 epochs of fine tuning with a learning rate of $0.01$ and a batch size of $128$.

Validation loss and accuracy were monitored during both training procedures.
The model state that enabled the greatest validation accuracy during training was restored for evaluations that followed.
Training data augmentations were limited to horizontal flipping (ImageNet and Tiny-ImageNet) and random cropping (ImageNet).

Training was performed using single NVIDIA 3090 and A100 GPUs.
Each model took approximately 12 hours to train on ImageNet and less than 30 minutes to fine-tune on Tiny-ImageNet.

\paragraph{Evaluating V1 Alignment}
We evaluated the similarity between neuro-constrained models of V1 and the primate primary visual cortex using the Brain-Score V1 benchmark \citep{schrimpf_brain-score_2020}.
The V1 benchmark score is an average of two sub-metrics: `V1 FreemanZiemba2013' and `V1 Marques2020', which we refer to as V1 Predictivity and V1 Property scores in what follows.
For each metric, the activity of artificial neurons in a given neural network layer is computed using in-silico neurophysiology experiments.
The V1 Predictivity score reflects the degree to which the model can explain the variance in stimulus-driven responses of V1 neurons, as determined by partial least squares regression mapping.
The V1 Property score measures how closely the distribution of $22$ different neural properties, from $7$ neural tuning categories (orientation, spatial frequency, response selectivity, receptive field size, surround modulation, texture modulation, and response magnitude), matches between the model's artificial neural responses and empirical data from macaque \acs{V1}.
Together, these two scores provide a comprehensive view of stimulus response similarity between of artificial and primate \acs{V1} neurons.

Brain-Score evaluations assume a defined mapping between units of an \acs{ANN} layer and a given brain region.
In all analyses of \acs{V1} alignment that follow, we systematically fixed the output neurons of ResNet50 layer 1 as the artificial V1 neurons.
Note that this is a stricter rule than most models submitted to the Brain-Score leaderboard, as researchers are able to choose which layer in the deep net should correspond to the V1 readout.
In baseline analyses, among multiple evaluated layers, we observed highest V1 alignment between artificial units primate \acs{V1} activity from layer 1, establishing it as a strong baseline.
Alternate layer \acs{V1} scores are presented in Appendix~\ref{app:alt-baselines}.

\section{Results}

\subsection{Architectural Components in Isolation}
\label{section:individual-components}

Patterns of neural activity observed in the brain can be attributed to the interplay of multiple specialized processes.
Through an isolated analysis, our initial investigations revealed the contribution of specialized mechanisms to explaining patterns of neural activity in \acrshort{V1}.
Tables \ref{tab:V1-comp-overview} and \ref{tab:V1-comp-properties} present the results of this analysis, including ImageNet validation accuracy, V1 Overall, V1 Predictivity, and V1 Property scores.

Among the four modules evaluated in this analysis, cortical magnification emerged as the most influential factor in enhancing V1 alignment.
This mechanism substantially improved the ResNet's ability to explain the variance in stimulus responses, and the artificial neurons exhibited tuning properties that were more closely aligned with those of biological neurons, particularly in terms of orientation tuning, spatial frequency tuning, response selectivity, and most of all, stimulus response magnitude.
However, the artificial neuronal responses of the cortical magnification network showed lower resemblance to those observed in primate V1 with regard to surround modulation, as compared to the baseline network.

Simulating neural normalization within the ResNet resulted in artificial neurons that displayed improved alignment with primate V1 in terms of response properties.
Noteworthy enhancements were observed in the spatial frequency, receptive field size, surround modulation, and response magnitude properties of neurons within the modified network, leading to improvements in the V1 Property score.
These results applied to both tuned and untuned forms of normalization.

In contrast, the introduction of center-surround convolutions yielded minimal improvements in neural predictivity and slight reductions in overall neuron property similarity.
Surprisingly, the surround modulation properties of the artificial neurons decreased compared to the baseline model, contrary to our expectations.

Finally, replacing the final $3 \times 3$ convolution preceding the artificial V1 readout with a locally connected layer resulted in modest changes in V1 alignment.
This was one of the two mechanisms that led to improvements in the surround modulation response property score (tuned normalization being the other).

These findings collectively provide valuable insights into the individual contributions of each specialized mechanism.
Although mechanisms simulating center-surround antagonism (i.e., \acs{DoG} convolution) and local connectivity provide little benefit to overall predictivity and property scores in isolation, we observed that they reduce the property dissimilarity gap among tuning properties that are nonetheless important and complement alignment scores where divisive normalization and cortical magnification do not.

\begin{table}[t]
    \centering
    \renewcommand{\arraystretch}{1.2}
    \resizebox{\linewidth}{!}{\begin{tabular}{lrrrr}
        & ImageNet Acc & V1 Overall & V1 Predictivity & V1 Property \\
        \hline
        ResNet50 (Baseline) & $.613 \pm .002$ & $.550 \pm .004$ & $.295 \pm .003$ & $.805 \pm .011$ \\
        Center-surround antagonism & \cellcolor{red!3}$.610 \pm .001$ & \cellcolor{red!5} $.545 \pm .002$ & \cellcolor{green!10} $.304 \pm .016$ & \cellcolor{red!20} $.786 \pm .018$ \\
        Local receptive fields & \cellcolor{red!4}$.609 \pm .001$ & $.550 \pm .006$ & \cellcolor{green!5} $.300 \pm .002$ & \cellcolor{red!5} $.799 \pm .012$ \\
        Divisive normalization & \cellcolor{red!7}$.606 \pm .001$ & \cellcolor{red!7}$.543 \pm .003$ & \cellcolor{red!24}$.271 \pm .014$ & \cellcolor{green!10}$.815 \pm .011$ \\
        Tuned normalization & \cellcolor{red!5}$.608 \pm .002$ & \cellcolor{red!3} $.547 \pm .004$ & \cellcolor{red!21} $.274 \pm .004$ & \cellcolor{green!15} $.820 \pm .009$ \\
        Cortical magnification & \cellcolor{red!65}$.548 \pm .008$ & \cellcolor{green!37} $.587 \pm .014$ & \cellcolor{green!75} $.370 \pm .008$ & $.805 \pm .021$ \\
    \end{tabular}}
    \caption{ImageNet object recognition classification performance ($64 \times 64$ images) and \acf{V1} alignment scores of ResNet50 augmented with each architectural component. Mean and standard deviations are reported across three runs (random initialization, training, and evaluating) of each architecture.  Scores higher than baseline are presented in green and those lower are presented with in red (the more saturated the color is, the greater the difference from baseline).}
    \label{tab:V1-comp-overview}
\end{table}

\begin{table}[t]
    \centering
    \renewcommand{\arraystretch}{1.2}
    \resizebox{\linewidth}{!}{\begin{tabular}{lrrrrrrr}
        &   & Spatial  & Response  &  & Surround  & Texture  & Response  \\
        &  Orientation & frequency &  selectivity & RF size &  modulation &  modulation &  magnitude \\
        \hline
        ResNet50 (Baseline) & $.893 \pm .023$ & $.826 \pm .048$ & $.684 \pm .059$ & $.832 \pm .080$ & $.820 \pm .009$ & $.786 \pm .058$ & $.790 \pm .042$ \\
        Center-surround & \cellcolor{red!17} $.876 \pm .027$ & \cellcolor{green!5} $.831 \pm .030$ & \cellcolor{red!52} $.632 \pm .012$ &\cellcolor{green!21} $.853 \pm .046$ & \cellcolor{red!47} $.773 \pm .027$ & \cellcolor{red!29} $.757 \pm .025$ & \cellcolor{red!7} $.783 \pm .024$ \\
        Local receptive fields & \cellcolor{green!11} $.904 \pm .021$ & \cellcolor{red!9} $.817 \pm .016$ & \cellcolor{red!36} $.648 \pm .008$ & \cellcolor{green!20} $.852 \pm .054$ & \cellcolor{green!27} $.847 \pm .083$ & \cellcolor{red!43} $.743 \pm .036$ & \cellcolor{red!10} $.780 \pm .022$ \\
        Divisive normalization & \cellcolor{green!15}$.908 \pm .017$ & \cellcolor{green!14}$.840 \pm .014$ & \cellcolor{green!5}$.689 \pm .007$ & \cellcolor{green!26}$.858 \pm .046$ & \cellcolor{green!40}$.860 \pm .070$ & \cellcolor{red!40}$.746 \pm .030$ & \cellcolor{green!56}$.846 \pm .019$ \\
        Tuned normalization  & \cellcolor{green!14} $.907 \pm .035$ & \cellcolor{green!15} $.841 \pm .013$ & \cellcolor{green!5} $.689 \pm .023$ & \cellcolor{green!33} $.865 \pm .031$ & \cellcolor{green!32} $.852 \pm .020$ & \cellcolor{red!43} $.742 \pm .029$ & \cellcolor{green!54} $.844 \pm .015$ \\
        Cortical magnification  & \cellcolor{green!14} $.907 \pm .037$ & \cellcolor{green!22} $.848 \pm .039$ & \cellcolor{green!24} $.708 \pm .011$ & \cellcolor{red!29} $.803 \pm .044$ & \cellcolor{red!100} $.664 \pm .063$ & \cellcolor{green!3} $.789 \pm .058$ & \cellcolor{green!100} $.917 \pm .071$ \\
    \end{tabular}}
    \caption{Model alignment across the seven \acf{V1} tuning properties that constitute the V1 Property score.  Mean and standard deviation of scores observed across three trials of model training and evaluation are reported.}
    \label{tab:V1-comp-properties}
\end{table}

\subsection{Complementary Components Explain V1 Activity}
\label{section:multiple-components}

Constraining a general-purpose deep learning model with a single architectural component is likely insufficient to explain primate \acrshort{V1} activity given our knowledge that a composition of known circuits play pivotal roles in visual processing.
Taking inspiration from this design principle, we supplemented a ResNet50 with each implemented architectural component and discern the necessary components to achieve optimal \acrshort{V1} alignment in an ablation study.
We omit the architectural component implementing divisive normalization, however, as it it cannot be integrated simultaneously with tuned normalization, which was observed to yield slightly higher V1 Predictivity and Property scores in isolated component evaluation.
In this study, we employed a greedy approach reminiscent of backward elimination feature selection.
In each round of this iterative approach, we selectively removed the architectural component that reduced overall \acs{V1} alignment the most until only one feature remained.
This analysis allowed us to identify the subset of components that collectively yielded the most significant improvements in V1 alignment, and unraveled the intricate relationship between these specialized features and their combined explanation of \acs{V1}.

The results of the ablation study are presented in Table \ref{tab:ablation}.
With the exception of center-surround antagonism, removing any neural mechanisms from the modified residual network reduced overall \acs{V1} alignment, suggesting that (1) each architectural component contributed to \acs{V1} alignment (the utility of center-surround antagonism is detailed in Section~\ref{section:adv-training}) and (2) nontrivial interactions between these mechanisms explain \acs{V1} more than what is possible with any single mechanism.
Seven of the eight models evaluated in this ablation study substantially outperformed all existing models on the Brain-Score platform in modeling \acs{V1} tuning property distributions.
Furthermore, four models were observed to achieve state-of-the-art V1 Overall scores, explaining both \acs{V1} stimulus response activity and neural response properties with high fidelity.

Whether or not feed-forward, ImageNet-trained \acp{ANN} can fully approximate activity in primate \acs{V1} has stood as an open question.
Previous studies have argued that no current model is capable of explaining all behavioral properties using neurons from a single readout layer \citep{marques_multi-scale_2021}.
The top performing models of the current evaluation stand out as the first examples of \acp{CNN} with neural representations that accurately approximate all evaluated \acs{V1} tuning properties (Appendix~\ref{app:tuning-props}), offering positive evidence for the efficacy of explaining primate \acs{V1} with neuro-inspired deep learning architectures.

\begin{table}[t]
    \centering
    \renewcommand{\arraystretch}{1.2}
    \resizebox{\linewidth}{!}{\begin{tabular}{ccccc|rrrr}
         Center- & Local & Tuned Nor- & Cortical Mag- & Adversarial & & \\
         Surround &  RF & malization & nification & Training & ImageNet Acc & V1 Overall & V1 Predictivity & V1 Property \\
        \hline
        \checkmark & \checkmark & \checkmark & \checkmark & & $.551$ & $.605$ & $.357$ & $.852$ \\
         & \checkmark & \checkmark & \checkmark & & $.543$ & $.605$ & $.353$ & $.857$ \\
        \checkmark &  & \checkmark & \checkmark & & $.541$ & $.599$ & $.340$ & $\mathbf{.858}$ \\
        \checkmark & \checkmark &  & \checkmark & & $.552$ & $.592$ & $.364$ & $.820$ \\
        \checkmark & \checkmark & \checkmark & & & $.603$ & $.555$ & $.276$ & $.834$ \\
         &  & \checkmark & \checkmark & & $.541$ & $.598$ & $.351$ & $.845$ \\
         & \checkmark &  & \checkmark & & $.555$ & $.593$ & $.384$ & $.803$ \\
         & \checkmark & \checkmark &  & &  $\mathbf{.606}$ & $.561$ & $.287$ & $.835$ \\
         \hline
         \checkmark & \checkmark & \checkmark & \checkmark & \checkmark & $.448$ & $\mathbf{.629}$ & $\mathbf{.430}$ & $.829$ \\
          & \checkmark & \checkmark & \checkmark & \checkmark & $.448$ & $.625$ & $\mathbf{.430}$ & $.819$ \\
    \end{tabular}}
    \caption{Backward component elimination results. Checkmarks denote whether or not the architectural component was included in the model.  Adversarial training was performed on the two models that tied for the top V1 Overall Score.}
    \label{tab:ablation}
\end{table}

\subsection{Object Recognition Robustness to Corrupted Images}

In contrast with the human visual system, typical \acp{CNN} generalize poorly to out-of-distribution data.
Small perturbations to an image can cause a model to output drastically different predictions than it would on the in-tact image.
Recent studies have demonstrated a positive correlation between model-brain similarity and robustness to image corruptions \citep{li_learning_2019,dapello_simulating_2020,reddy_biologically_2020,safarani_towards_2021,cirincione_implementing_nodate}
After freezing the brain-aligned representations of the models presented in this work and fine-tuning each model's classification head on Tiny-ImageNet (see Section~\ref{subsec:training-evaluation}), we evaluated the object recognition accuracy of each model from Section~\ref{section:individual-components} and the top two overall models from section \ref{section:multiple-components} on the Tiny-ImageNet-C dataset.
The results of these evaluations for each category of corruption and corruption strength are provided in Appendix~\ref{app:tinycorruptions}.

Among the evaluated components, only tuned normalization was observed to yield improved corrupt image classification accuracy over the entire test set, albeit slight, beating the baseline accuracy ($0.278$) by $0.005$ (i.e., an absolute improvement of $.5\%$).
More substantial improvements were observed on `brightness', `defocus\_blur', `elastic\_transform', and `pixelate' corruptions (improvements over the baseline of .00986, .00989, .0105, and .0133, respectively).

\subsection{Adversarially Training Neuro-Constrained ResNets}
\label{section:adv-training}

Adversarial training has previously been shown to enhance the brain-similarity of artificial neural representations without any modification to the underlying network \citep{dapello_simulating_2020,riedel_bag_2022}.
Curious as to whether adversarial training would further align the neuro-constrained ResNet50s with \acs{V1} activity, we selectively trained the two networks most aligned with \acs{V1} (one model with all architectural components and the other with all components except center-surround convolution) from Section~\ref{section:multiple-components} using ``Free'' adversarial training \citep{shafahi_adversarial_2019}.
Despite the drop in object recognition accuracy, the artificial neural representations that emerged in each network were drastically better predictors of stimulus response variance representations.
Tuning property alignment dropped in the process, but remained above previous state-of-the-art regardless.
Interestingly, we found that the main difference in \acs{V1} scores between these two models can be traced to surround modulation tuning alignment.
Center-surround convolution indeed contributed to improved surround modulation tuning learned while training with on corrupted images, contrasting its apparent lack of contribution to the overall network suggested in the ablation study. 

In sum, both networks achieved Rank-1 V1 Overall, Predictivity, and Property scores by large margins, setting a new standard in this breed of brain-aligned \acp{CNN}.
At the time of writing, the previous Rank-1 V1 Overall, Predictivity, and Property scores were .594, .409, and .816, respectively, and all achieved by separate models.

\section{Discussion}

Throughout this work we presented a systematic evaluation of four architectural components derived from neuroscience principles and their influence on model-V1 similarity.
Specifically, we developed novel \acs{ANN} layers that simulate principle processing mechanisms of the primate visual stream including center-surround antagonism, local receptive fields, tuned normalization, and cortical magnification.
Through an ablation study and isolated component analyses, we found that each component contributed to the production of latent \acs{ANN} representations that better resemble those of primate \acs{V1}, as compared to a traditional baseline \acs{CNN}.
When these four components were assembled together within a neuro-constrained ResNet50, \acs{V1} tuning properties were explained better than any previous deep learning model that we are aware of.
Furthermore, this neuro-constrained model exhibited state-of-the-art explanation of V1 neural activity and is the first of its kind to do so, by a large margin nonetheless, highlighting a promising direction in biologically constrained \acp{ANN}.
Training this model with ``free'' adversarial training greatly improved its ability to predict primate neural response to image stimuli at a minor sacrifice to tuning property similarity, establishing an even larger gap between previous state of the art.

Among all architectural components examined in this work, cortical magnification was the most influential to improving \acs{V1} alignment.
This mechanism on its own could not explain the neural activity as completely as the top models of this study, however.
Our implementation of tuned normalization provided substantial improvement to \acs{V1} tuning property alignment, and was the only component that contributed to model robustness.
The importance of center-surround antagonism seemed to be training data-dependent.
In our ablation study, for which all models were trained on ImageNet, center-surround convolutional layers did not contribute to overall \acs{V1} scores.
This did not surprise us, as deep \acp{CNN} have the capacity to learn similar representations without these specialized layers.
When training on adversarially perturbed data, however, the center-surround antagonism provided by this layer appeared to improve surround modulation tuning properties of artificial V1 neurons.
While previous attempts at improving model-brain similarity have been highly dataset dependent, our results highlight the importance of artificial network design.

A notable limitation to our work is the reduction in ImageNet classification performance that was observed upon the introduction of cortical magnification.
While perfectly maintaining baseline model accuracy was not a motivation of this work, we can imagine situations in which object recognition performance needs to be preserved alongside these improvements in brain-model alignment.
One scope of future work involves implementing saliency-driven polar transformations, so that the center of the polar map assumed by the polar transform is focused on an object of interest as opposed to being fixed at the center of the image.
We expect that such a mechanism would help to mitigate these reductions in ImageNet accuracy.

We additionally plan to extend this work to model architectures other than ResNet to validate the widespread application of each of these components.

This work highlights an important advancement in the field of NeuroAI, as we systematically establish a set of neuro-constrained architectural components that contribute to state-of-the-art \acs{V1} alignment.
We argue that our architecture-driven approach can be further generalized to additional areas of the brain as well.
The neuroscience insights that could be gleaned from increasingly accurate in-silico models of the brain have the potential to transform the fields of both neuroscience and \acs{AI}.

\newpage

\newpage
{\small
\bibliographystyle{unsrt}
\bibliography{2023-NeurIPS-V1}

\begin{thebibliography}{10}

\bibitem{carandini_we_2005}
Matteo Carandini, Jonathan~B. Demb, Valerio Mante, David~J. Tolhurst, Yang Dan,
  Bruno~A. Olshausen, Jack~L. Gallant, and Nicole~C. Rust.
\newblock Do {We} {Know} {What} the {Early} {Visual} {System} {Does}?
\newblock {\em Journal of Neuroscience}, 25(46):10577--10597, November 2005.

\bibitem{yamins_hierarchical_2013}
Daniel~L Yamins, Ha~Hong, Charles Cadieu, and James~J DiCarlo.
\newblock Hierarchical {Modular} {Optimization} of {Convolutional} {Networks}
  {Achieves} {Representations} {Similar} to {Macaque} {IT} and {Human}
  {Ventral} {Stream}.
\newblock In {\em Advances in {Neural} {Information} {Processing} {Systems}},
  volume~26. Curran Associates, Inc., 2013.

\bibitem{yamins_performance-optimized_2014}
Daniel L.~K. Yamins, Ha~Hong, Charles~F. Cadieu, Ethan~A. Solomon, Darren
  Seibert, and James~J. DiCarlo.
\newblock Performance-optimized hierarchical models predict neural responses in
  higher visual cortex.
\newblock {\em Proceedings of the National Academy of Sciences},
  111(23):8619--8624, June 2014.

\bibitem{yamins_using_2016}
Daniel L.~K. Yamins and James~J. DiCarlo.
\newblock Using goal-driven deep learning models to understand sensory cortex.
\newblock {\em Nature Neuroscience}, 19(3):356--365, March 2016.
\newblock Number: 3 Publisher: Nature Publishing Group.

\bibitem{majaj_simple_2015}
Najib~J. Majaj, Ha~Hong, Ethan~A. Solomon, and James~J. DiCarlo.
\newblock Simple {Learned} {Weighted} {Sums} of {Inferior} {Temporal}
  {Neuronal} {Firing} {Rates} {Accurately} {Predict} {Human} {Core} {Object}
  {Recognition} {Performance}.
\newblock {\em Journal of Neuroscience}, 35(39):13402--13418, September 2015.
\newblock Publisher: Society for Neuroscience Section: Articles.

\bibitem{cadena_deep_2019}
Santiago~A. Cadena, George~H. Denfield, Edgar~Y. Walker, Leon~A. Gatys,
  Andreas~S. Tolias, Matthias Bethge, and Alexander~S. Ecker.
\newblock Deep convolutional models improve predictions of macaque {V1}
  responses to natural images.
\newblock {\em PLOS Computational Biology}, 15(4):e1006897, April 2019.
\newblock Publisher: Public Library of Science.

\bibitem{bashivan_neural_2019}
Pouya Bashivan, Kohitij Kar, and James~J. DiCarlo.
\newblock Neural population control via deep image synthesis.
\newblock {\em Science}, 364(6439):eaav9436, May 2019.
\newblock Publisher: American Association for the Advancement of Science.

\bibitem{hubel_receptive_1962}
David~H Hubel and Torsten~N Wiesel.
\newblock Receptive fields, binocular interaction and functional architecture
  in the cat's visual cortex.
\newblock {\em The Journal of physiology}, 160(1):106, 1962.
\newblock Publisher: Wiley-Blackwell.

\bibitem{schrimpf_brain-score_2020}
Martin Schrimpf, Jonas Kubilius, Ha~Hong, Najib~J. Majaj, Rishi Rajalingham,
  Elias~B. Issa, Kohitij Kar, Pouya Bashivan, Jonathan Prescott-Roy, Franziska
  Geiger, Kailyn Schmidt, Daniel L.~K. Yamins, and James~J. DiCarlo.
\newblock Brain-{Score}: {Which} {Artificial} {Neural} {Network} for {Object}
  {Recognition} is most {Brain}-{Like}?, January 2020.
\newblock Pages: 407007 Section: New Results.

\bibitem{he_deep_2016}
Kaiming He, Xiangyu Zhang, Shaoqing Ren, and Jian Sun.
\newblock Deep residual learning for image recognition.
\newblock In {\em Proceedings of the IEEE conference on computer vision and
  pattern recognition}, pages 770--778, 2016.

\bibitem{tan_efficientnet_2019}
Mingxing Tan and Quoc Le.
\newblock Efficientnet: Rethinking model scaling for convolutional neural
  networks.
\newblock In {\em International conference on machine learning}, pages
  6105--6114. PMLR, 2019.

\bibitem{li_learning_2019}
Zhe Li, Wieland Brendel, Edgar Walker, Erick Cobos, Taliah Muhammad, Jacob
  Reimer, Matthias Bethge, Fabian Sinz, Zachary Pitkow, and Andreas Tolias.
\newblock Learning from brains how to regularize machines.
\newblock In {\em Advances in {Neural} {Information} {Processing} {Systems}},
  volume~32. Curran Associates, Inc., 2019.

\bibitem{dapello_simulating_2020}
Joel Dapello, Tiago Marques, Martin Schrimpf, Franziska Geiger, David Cox, and
  James~J DiCarlo.
\newblock Simulating a {Primary} {Visual} {Cortex} at the {Front} of {CNNs}
  {Improves} {Robustness} to {Image} {Perturbations}.
\newblock In {\em Advances in {Neural} {Information} {Processing} {Systems}},
  volume~33, pages 13073--13087. Curran Associates, Inc., 2020.

\bibitem{reddy_biologically_2020}
Manish~V. Reddy, Andrzej Banburski, Nishka Pant, and Tomaso Poggio.
\newblock Biologically {Inspired} {Mechanisms} for {Adversarial} {Robustness},
  June 2020.
\newblock arXiv:2006.16427 [cs, stat].

\bibitem{allman_stimulus_1985}
J.~Allman, F.~Miezin, and E.~McGuinness.
\newblock Stimulus specific responses from beyond the classical receptive
  field: neurophysiological mechanisms for local-global comparisons in visual
  neurons.
\newblock {\em Annual Review of Neuroscience}, 8:407--430, 1985.

\bibitem{knierim_neuronal_1992}
J.~J. Knierim and D.~C. van Essen.
\newblock Neuronal responses to static texture patterns in area {V1} of the
  alert macaque monkey.
\newblock {\em Journal of Neurophysiology}, 67(4):961--980, April 1992.
\newblock Publisher: American Physiological Society.

\bibitem{walker_asymmetric_1999}
Gary~A. Walker, Izumi Ohzawa, and Ralph~D. Freeman.
\newblock Asymmetric {Suppression} {Outside} the {Classical} {Receptive}
  {Field} of the {Visual} {Cortex}.
\newblock {\em Journal of Neuroscience}, 19(23):10536--10553, December 1999.
\newblock Publisher: Society for Neuroscience Section: ARTICLE.

\bibitem{shen_cue-invariant_2007}
Zhi-Ming Shen, Wei-Feng Xu, and Chao-Yi Li.
\newblock Cue-invariant detection of centre–surround discontinuity by {V1}
  neurons in awake macaque monkey.
\newblock {\em The Journal of Physiology}, 583(Pt 2):581--592, September 2007.

\bibitem{deangelis_length_1994}
Gregory~C DeAngelis, RALPH~D Freeman, and IZUMI Ohzawa.
\newblock Length and width tuning of neurons in the cat's primary visual
  cortex.
\newblock {\em Journal of neurophysiology}, 71(1):347--374, 1994.

\bibitem{sceniak_contrasts_1999}
Michael~P Sceniak, Dario~L Ringach, Michael~J Hawken, and Robert Shapley.
\newblock Contrast's effect on spatial summation by macaque {V1} neurons.
\newblock {\em Nature neuroscience}, 2(8):733--739, 1999.
\newblock Publisher: Nature Publishing Group.

\bibitem{sceniak_visual_2001}
Michael~P Sceniak, Michael~J Hawken, and Robert Shapley.
\newblock Visual spatial characterization of macaque {V1} neurons.
\newblock {\em Journal of neurophysiology}, 85(5):1873--1887, 2001.
\newblock Publisher: American Physiological Society Bethesda, MD.

\bibitem{hasani_surround_2019}
Hosein Hasani, Mahdieh Soleymani, and Hamid Aghajan.
\newblock Surround {Modulation}: {A} {Bio}-inspired {Connectivity} {Structure}
  for {Convolutional} {Neural} {Networks}.
\newblock In H.~Wallach, H.~Larochelle, A.~Beygelzimer, F.~d' Alché-Buc,
  E.~Fox, and R.~Garnett, editors, {\em Advances in {Neural} {Information}
  {Processing} {Systems}}, volume~32. Curran Associates, Inc., 2019.

\bibitem{babaiee_-off_2021}
Zahra Babaiee, Ramin Hasani, Mathias Lechner, Daniela Rus, and Radu Grosu.
\newblock On-off center-surround receptive fields for accurate and robust image
  classification.
\newblock In {\em International {Conference} on {Machine} {Learning}}, pages
  478--489. PMLR, 2021.

\bibitem{koulakov_orientation_2001}
Alexei~A Koulakov and Dmitri~B Chklovskii.
\newblock Orientation preference patterns in mammalian visual cortex: a wire
  length minimization approach.
\newblock {\em Neuron}, 29(2):519--527, 2001.
\newblock Publisher: Elsevier.

\bibitem{lee_topographic_2020}
Hyodong Lee, Eshed Margalit, Kamila~M. Jozwik, Michael~A. Cohen, Nancy
  Kanwisher, Daniel L.~K. Yamins, and James~J. DiCarlo.
\newblock Topographic deep artificial neural networks reproduce the hallmarks
  of the primate inferior temporal cortex face processing network.
\newblock preprint, Neuroscience, July 2020.

\bibitem{pogodin_towards_2021}
Roman Pogodin, Yash Mehta, Timothy Lillicrap, and Peter~E Latham.
\newblock Towards {Biologically} {Plausible} {Convolutional} {Networks}.
\newblock In {\em Advances in {Neural} {Information} {Processing} {Systems}},
  volume~34, pages 13924--13936. Curran Associates, Inc., 2021.

\bibitem{carandini_normalization_2012}
Matteo Carandini and David~J. Heeger.
\newblock Normalization as a canonical neural computation.
\newblock {\em Nature Reviews Neuroscience}, 13(1):51--62, January 2012.
\newblock Number: 1 Publisher: Nature Publishing Group.

\bibitem{heeger_recurrent_2020}
David~J. Heeger and Klavdia~O. Zemlianova.
\newblock A recurrent circuit implements normalization, simulating the dynamics
  of {V1} activity.
\newblock {\em Proceedings of the National Academy of Sciences},
  117(36):22494--22505, September 2020.
\newblock Publisher: Proceedings of the National Academy of Sciences.

\bibitem{burg_learning_2021}
Max~F Burg, Santiago~A Cadena, George~H Denfield, Edgar~Y Walker, Andreas~S
  Tolias, Matthias Bethge, and Alexander~S Ecker.
\newblock Learning divisive normalization in primary visual cortex.
\newblock {\em PLOS Computational Biology}, 17(6):e1009028, 2021.
\newblock Publisher: Public Library of Science San Francisco, CA USA.

\bibitem{cirincione_implementing_nodate}
Andrew Cirincione, Reginald Verrier, Artiom Bic, Stephanie Olaiya, James~J
  DiCarlo, Lawrence Udeigwe, and Tiago Marques.
\newblock Implementing {Divisive} {Normalization} in {CNNs} {Improves}
  {Robustness} to {Common} {Image} {Corruptions}.

\bibitem{rust_how_2006}
Nicole~C. Rust, Valerio Mante, Eero~P. Simoncelli, and J.~Anthony Movshon.
\newblock How {MT} cells analyze the motion of visual patterns.
\newblock {\em Nature Neuroscience}, 9(11):1421--1431, November 2006.
\newblock Number: 11 Publisher: Nature Publishing Group.

\bibitem{morrone_functional_1982}
M~Concetta Morrone, DC~Burr, and Lamberto Maffei.
\newblock Functional implications of cross-orientation inhibition of cortical
  visual cells. {I}. {Neurophysiological} evidence.
\newblock {\em Proceedings of the Royal Society of London. Series B. Biological
  Sciences}, 216(1204):335--354, 1982.
\newblock Publisher: The Royal Society London.

\bibitem{deangelis_organization_1992}
GC~DeAngelis, JG~Robson, I~Ohzawa, and RD~Freeman.
\newblock Organization of suppression in receptive fields of neurons in cat
  visual cortex.
\newblock {\em Journal of Neurophysiology}, 68(1):144--163, 1992.

\bibitem{daniel_representation_1961}
PM~Daniel and D~Whitteridge.
\newblock The representation of the visual field on the cerebral cortex in
  monkeys.
\newblock {\em The Journal of physiology}, 159(2):203, 1961.
\newblock Publisher: Wiley-Blackwell.

\bibitem{schwartz_spatial_1977}
Eric~L Schwartz.
\newblock Spatial mapping in the primate sensory projection: analytic structure
  and relevance to perception.
\newblock {\em Biological cybernetics}, 25(4):181--194, 1977.
\newblock Publisher: Springer.

\bibitem{schwartz_computational_1980}
Eric~L Schwartz.
\newblock Computational anatomy and functional architecture of striate cortex:
  a spatial mapping approach to perceptual coding.
\newblock {\em Vision research}, 20(8):645--669, 1980.
\newblock Publisher: Elsevier.

\bibitem{schwartz_computational_1994}
Eric~L Schwartz.
\newblock Computational studies of the spatial architecture of primate visual
  cortex: columns, maps, and protomaps.
\newblock {\em Primary visual cortex in primates}, pages 359--411, 1994.
\newblock Publisher: Springer.

\bibitem{polimeni_multi-area_2006}
Jonathan~R Polimeni, Mukund Balasubramanian, and Eric~L Schwartz.
\newblock Multi-area visuotopic map complexes in macaque striate and
  extra-striate cortex.
\newblock {\em Vision research}, 46(20):3336--3359, 2006.
\newblock Publisher: Elsevier.

\bibitem{carlos_esteves_polar_2018}
Carlos Esteves, Christine Allen-Blanchette, Xiaowei Zhou, and Kostas
  Daniilidis.
\newblock Polar {Transformer} {Networks}.
\newblock {\em International Conference on Learning Representations}, 2018.

\bibitem{henriques_warped_2021}
João~F. Henriques and Andrea Vedaldi.
\newblock Warped {Convolutions}: {Efficient} {Invariance} to {Spatial}
  {Transformations}, 2021.
\newblock \_eprint: 1609.04382.

\bibitem{kiritani_recurrent_2020}
Taro Kiritani and Koji Ono.
\newblock Recurrent {Attention} {Model} with {Log}-{Polar} {Mapping} is
  {Robust} against {Adversarial} {Attacks}, 2020.
\newblock \_eprint: 2002.05388.

\bibitem{blackburn_simple_nodate}
M.R. Blackburn.
\newblock A {Simple} {Computational} {Model} of {Center}-{Surround} {Receptive}
  {Fields} in the {Retina}.
\newblock Technical report.
\newblock Section: Technical Reports.

\bibitem{deng_imagenet_2009}
Jia Deng, Wei Dong, Richard Socher, Li-Jia Li, Kai Li, and Li~Fei-Fei.
\newblock {ImageNet}: {A} large-scale hierarchical image database.
\newblock In {\em 2009 {IEEE} {Conference} on {Computer} {Vision} and {Pattern}
  {Recognition}}, pages 248--255, June 2009.
\newblock ISSN: 1063-6919.

\bibitem{hendrycks_benchmarking_2019}
Dan Hendrycks and Thomas Dietterich.
\newblock Benchmarking {Neural} {Network} {Robustness} to {Common}
  {Corruptions} and {Perturbations}, March 2019.
\newblock arXiv:1903.12261 [cs, stat].

\bibitem{marques_multi-scale_2021}
Tiago Marques, Martin Schrimpf, and James~J. DiCarlo.
\newblock Multi-scale hierarchical neural network models that bridge from
  single neurons in the primate primary visual cortex to object recognition
  behavior, August 2021.
\newblock Pages: 2021.03.01.433495 Section: New Results.

\bibitem{safarani_towards_2021}
Shahd Safarani, Arne Nix, Konstantin Willeke, Santiago Cadena, Kelli Restivo,
  George Denfield, Andreas Tolias, and Fabian Sinz.
\newblock Towards robust vision by multi-task learning on monkey visual cortex.
\newblock In {\em Advances in {Neural} {Information} {Processing} {Systems}},
  volume~34, pages 739--751. Curran Associates, Inc., 2021.

\bibitem{riedel_bag_2022}
Alexander Riedel.
\newblock Bag of {Tricks} for {Training} {Brain}-{Like} {Deep} {Neural}
  {Networks}.
\newblock March 2022.

\bibitem{shafahi_adversarial_2019}
Ali Shafahi, Mahyar Najibi, Mohammad~Amin Ghiasi, Zheng Xu, John Dickerson,
  Christoph Studer, Larry~S Davis, Gavin Taylor, and Tom Goldstein.
\newblock Adversarial training for free!
\newblock In {\em Advances in {Neural} {Information} {Processing} {Systems}},
  volume~32. Curran Associates, Inc., 2019.

\bibitem{madry_towards_2017}
Aleksander Madry, Aleksandar Makelov, Ludwig Schmidt, Dimitris Tsipras, and
  Adrian Vladu.
\newblock Towards deep learning models resistant to adversarial attacks.
\newblock {\em arXiv preprint arXiv:1706.06083}, 2017.

\end{thebibliography}
}

\newpage
\section*{Appendix}
\appendix
\counterwithin{figure}{section}

\section{Supplemental Model Diagrams}
\label{app:diagrams}

\begin{figure}[h!]
    \centering
    \includegraphics[width=\linewidth]{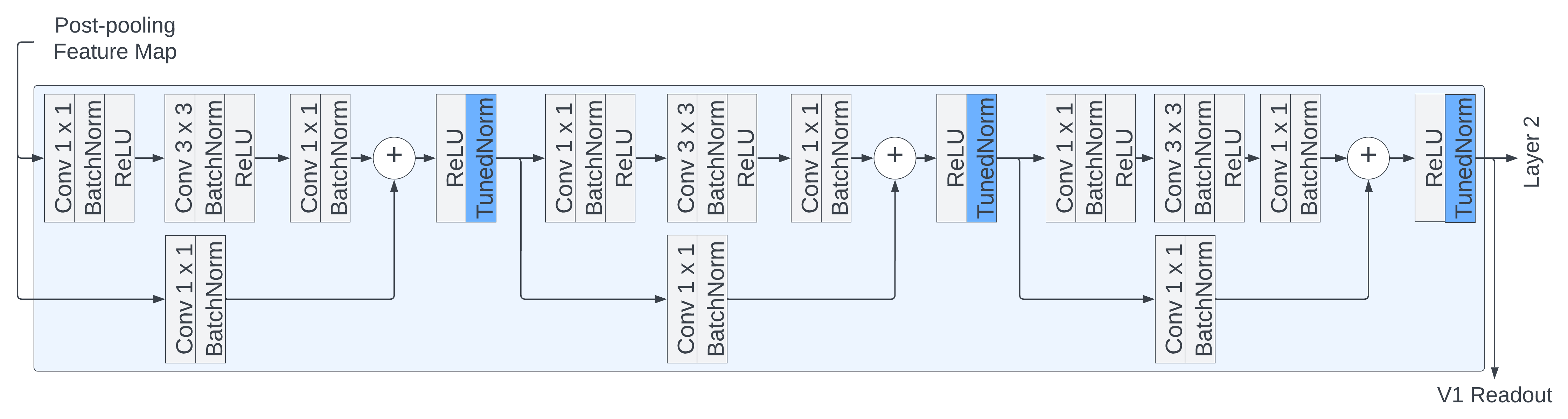}
    (E) Tuned normalization
    \includegraphics[width=\linewidth]{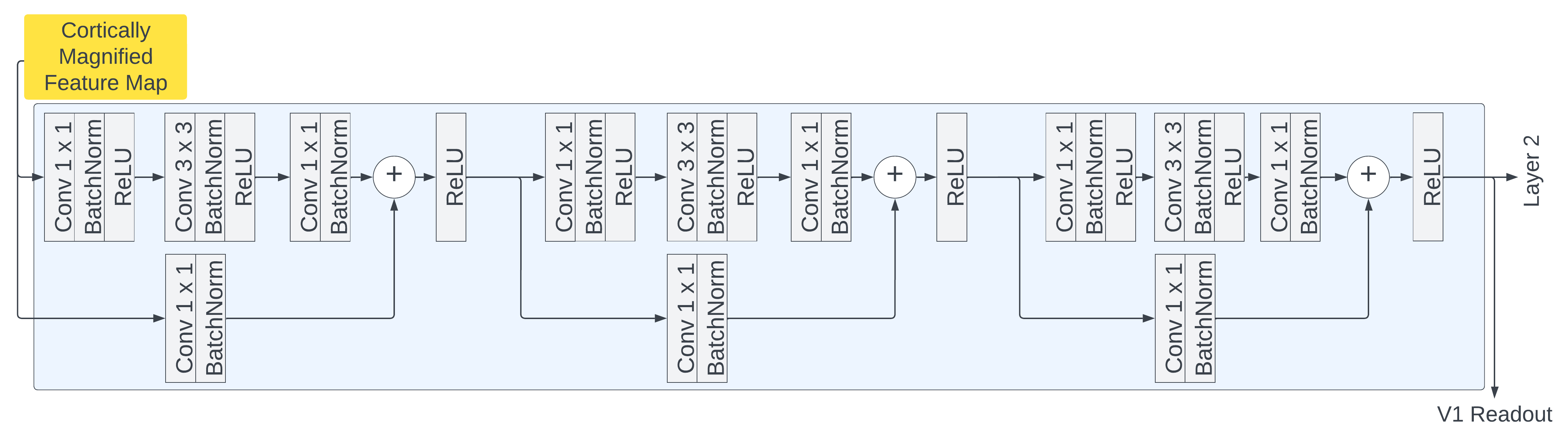}
    (F) Cortical magnification
    \caption{ResNet50 residual layer 1, supplemented with individual neuro-constrained architectural components, as in section \ref{section:individual-components}. (A) No modification (baseline ResNet50 layer 1), (B) with center-surround antagonism, (C) with local \acf{RF}, (D) with divisive normalization, (E) with tuned normalization, (F) with cortical magnification.}
    \label{fig:layer_isolated_2}
\end{figure}

Fig.~\ref{fig:layer_isolated_2} depicts the modifications made to ResNet50 residual layer 1 in the isolated component analyses of section \ref{section:individual-components}.
All multi-component (composite) models analyzed in Section~\ref{section:multiple-components} relied on combinations of these modifications (as exemplified in Fig.~\ref{fig:architecture}).

\begin{figure}[h!]
    \centering
    \includegraphics[width=\linewidth]{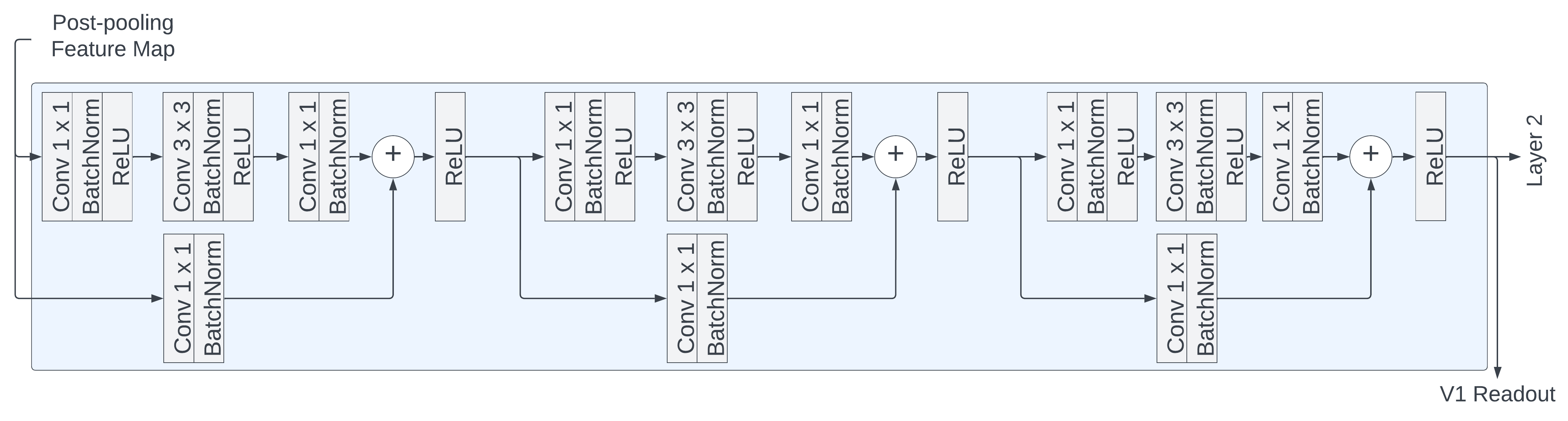}
    (A) ResNet50 baseline
    \includegraphics[width=\linewidth]{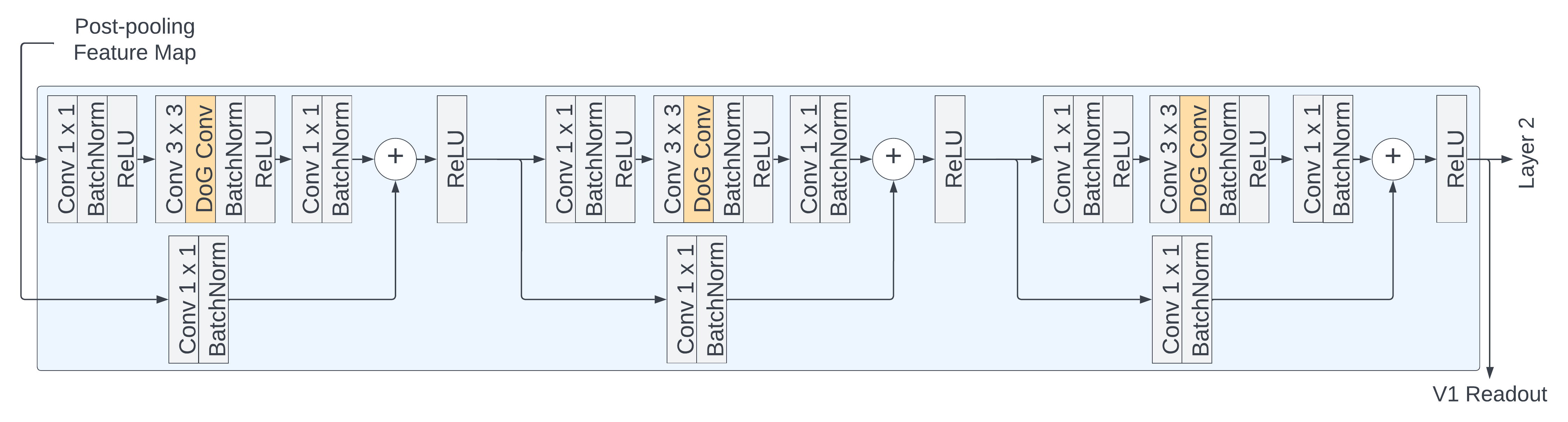}
    (B) Center-surround antagonism
    \includegraphics[width=\linewidth]{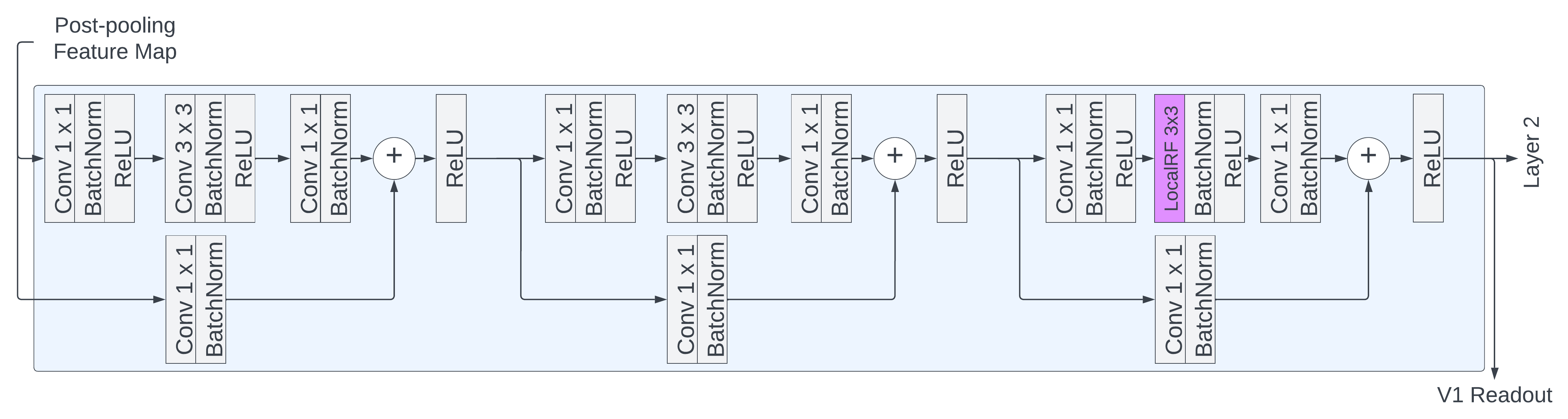}
    (C) Local receptive fields
    \includegraphics[width=\linewidth]{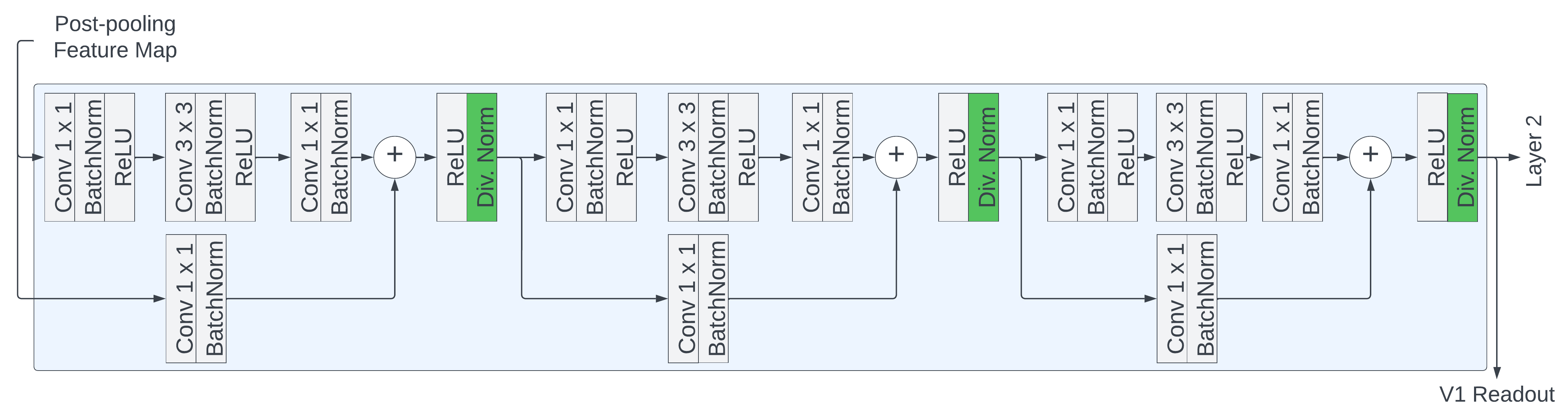}
    (D) Divisive normalization
    \label{fig:layer_isolated_1}
\end{figure}

\newpage

\newpage
\section{V1 Scores of Alternate Layers of Baseline Network}
\label{app:alt-baselines}

When evaluating a model on Brain-Score, users are permitted to commit a mapping between model layers and areas of the visual stream.
Model-brain alignment is computed for each mapped pair in the Brain-Score evaluation.
To promote a fair evaluation, we sought to find the layer that yielded optimal V1 alignment from the baseline ResNet50 model and fix this layer as the artificial V1 readout layer in all of our tested models. 
It is worth noting that after supplementing the base ResNet50 with neuro-constrained components, this layer may no longer offer optimal V1 alignment in the augmented network.
In spite of this, we maintain this layer as our artificial V1 readout layer for fair evaluation.

\begin{figure}[h!]
    \centering
    \includegraphics[width=\linewidth]{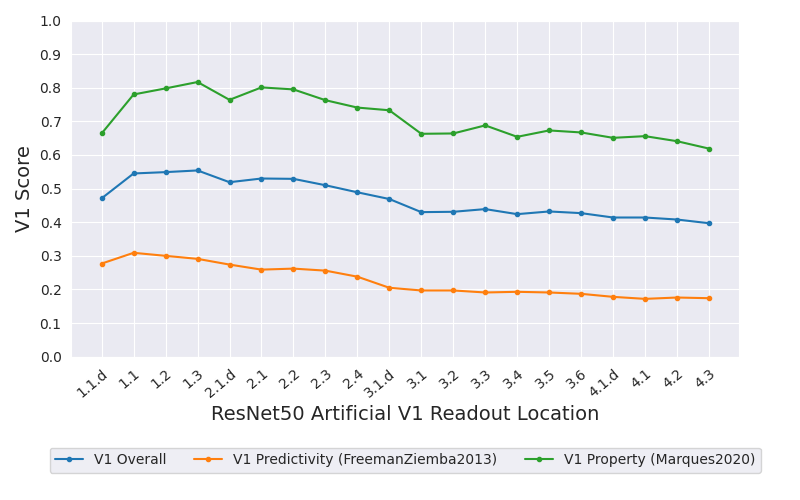}
    \caption{V1 alignment Brain-Scores for 20 different hidden layers of ResNet50.  In the plot above, readout location `X.Y' denotes that artificial V1 activity was evaluated from residual block `Y' of residual layer `X'.  Readout location suffixed with `.d' correspond to downsampling layers of the associated residual bottleneck.  Highest V1 overall score came from block 3 of residual layer 1.}
    \label{fig:base_layer_scores}
\end{figure}

To find the ResNet50 layer with the best V1 Overall, Predictivity, and Property scores, we compared a total of 20 different hidden layers (Fig.~\ref{fig:base_layer_scores}).
$16$ of these layers corresponded to the post-activation hidden states of the network.
The remaining $4$ were downsampling layers of the first bottleneck block of each residual layer in the network, as these have previously demonstrated good V1 alignment \citep{dapello_simulating_2020}.
Aside from these downsampling layers, hidden layers that did not follow a ReLU activation  were omitted from this evaluation as the activities of these states can take on negative values and are therefore less interpretable as neural activities.
Among all evaluated layers, the final output of ResNet50 residual layer 1 (i.e., the output of the third residual block of ResNet50) offered the highest V1 Overall score, and was therefore selected as the artificial V1 readout layer in all of our experiments.

\newpage
\section{Expanded Model Tuning Properties}
\label{app:tuning-props}

\Acf{V1} tuning property alignments for each composite model evaluated in Section~\ref{section:multiple-components} are presented in Table~\ref{tab:ablation-properties}.
Tuning property similarities are computed as ceiled Kolmogorov-Smirnov distance between artificial neural response distributions from the model and empirical distributions recorded in primates \citep{schrimpf_brain-score_2020, marques_multi-scale_2021}.

\begin{table}[h]
    \centering
    \setlength{\tabcolsep}{10pt} 
    \renewcommand{\arraystretch}{1.2}
    \resizebox{\linewidth}{!}{\begin{tabular}{ccccc|ccccccccc}
         \multicolumn{1}{c}{\rot[60]{Center-Surround}} & \multicolumn{1}{c}{\rot[60]{Local RF}} & \multicolumn{1}{c}{\rot[60]{Tuned Norm.}} & \multicolumn{1}{c}{\rot[60]{Cortical Mag.}} & \multicolumn{1}{c}{\rot[60]{Adv. Training}} & \rot[60]{Orientation} & \rot[60]{Spatial Frequency} & \rot[60]{Response Selectivity} & \rot[60]{RF Size} & \rot[60]{Surround Modulation} & \rot[60]{Texture Modulation} & \rot[60]{Response Mag.} & \\

        \hline
        \checkmark & \checkmark & \checkmark & \checkmark & & $.891$ & $.925$ & $.756$ & $.840$ & $.779$ & $.844$ & $.930$ \\
         & \checkmark & \checkmark & \checkmark & & $.858$ & $.919$ & $.780$ & $.834$ & $.808$ & $.871$ & $.930$ \\
        \checkmark &  & \checkmark & \checkmark & & $.894$ & $.932$ & $.750$ & $.851$ & $.775$ & $.858$ & $.946$ \\
        \checkmark & \checkmark &  & \checkmark & & $.878$ & $.873$ & $.739$ & $.816$ & $.719$ & $.802$ & $.910$ \\
        \checkmark & \checkmark & \checkmark & & & $.875$ & $.873$ & $.702$ & $.808$ & $.890$ & $.815$ & $.870$ \\
         &  & \checkmark & \checkmark & & $.873$ & $.886$ & $.735$ & $.840$ & $.794$ & $.825$ & $.959$ \\
         & \checkmark &  & \checkmark & & $.902$ & $.866$ & $.715$ & $.801$ & $.625$ & $.841$ & $.869$ \\
         & \checkmark & \checkmark &  & & $.915$ & $.817$ & $.691$ & $.811$ & $.898$ & $.802$ & $.911$ \\
         \hline
         \checkmark & \checkmark & \checkmark & \checkmark & \checkmark & $.924$ & $.863$ & $.773$ & $.797$ & $.733$ & $.815$ & $.899$ \\
          & \checkmark & \checkmark & \checkmark & \checkmark & $.944$ & $.834$ & $.768$ & $.806$ & $.673$ & $.811$ & $.900$ \\
    \end{tabular}}
    \caption{Ablation study model alignment across the seven \acf{V1} tuning properties that constitute the V1 Property score (`Marques2020') of Brain-Score. Checkmarks denote whether or not the architectural component was included in the model.}
    \label{tab:ablation-properties}
\end{table}

\newpage
\section{Adversarial Training}
\label{app:adv-training}

The neuro-constrained ResNets discussed in Section~\ref{section:adv-training} were trained using the ``Free'' adversarial training method proposed by Shafahi \emph{et al.}~\citep{shafahi_adversarial_2019}.
In Projected Gradient Descent (PGD)-based adversarial training (a typical approach to adversarially training robust classifiers) \citep{madry_towards_2017}, a network is trained on adversarial samples that are generated on the fly during training.
Specifically, in PGD-based adversarial training, a batch of adversarial images is first generated through a series of iterative perturbations to an original image batch, at which point the parameters of the network are finally updated according to the network's loss, as evaluated on the adversarial examples.
``Free'' adversarial training generates adversarial training images with a similar approach, but the parameters of the network are simultaneously updated with every iteration of image perturbation, significantly reducing training time.
The authors refer to these mini-batch updates as ``replays'', and refer to the number of replays of each mini-batch with the parameter $m$.

The adversarially trained models of Section~ref{section:adv-training} were trained with $m=4$ replays and perturbation clipping of $\epsilon = \frac{2}{255}$.
These models were trained for $120$ epochs using a stochastic gradient descent optimizer with an initial learning rate of $0.1$, which was reduced by a factor of $10$ every $40$ epochs, momentum of $0.9$, and weight decay of $1\times 10^{-5}$.
Each model was initialized with the weights that were learned during traditional ImageNet training for the analyses in Section~\ref{section:multiple-components}.
``Free'' adversarial training was performed using code provided by the authors of this method (\url{https://github.com/mahyarnajibi/FreeAdversarialTraining}).

\newpage
\section{Robustness to Common Image Corruptions}
\label{app:tinycorruptions}

\subsection{Dataset Description}
We evaluated image classification robustness to common image corruptions using the Tiny-ImageNet-C dataset \citep{hendrycks_benchmarking_2019}.
Recall that Tiny-ImageNet-C was used instead of ImageNet-C, because our models were trained on $64\times64$ input images. Downscaling ImageNet-C images would have potentially altered the intended corruptions and biased our evaluations.

\begin{figure}[h!]
    \centering
    \includegraphics[width=\linewidth]{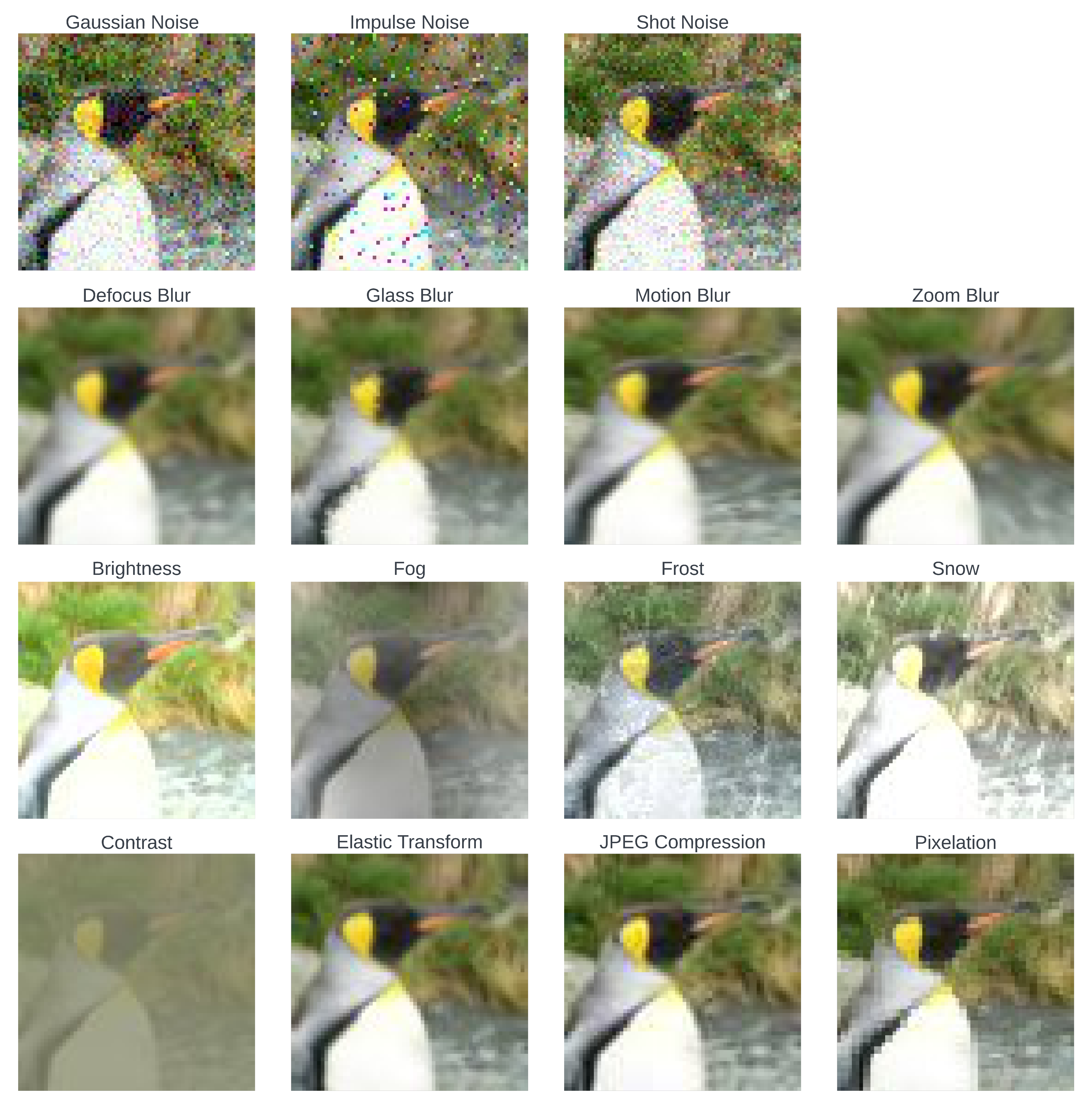}
    \caption{$15$ corruptions of the Tiny-ImageNet-C dataset, applied to a sample image from Tiny-ImageNet-C.  First row: noise corruptions, second row: blur corruptions, third row: weather corruptions, bottom row: digital corruptions.  All corruptions shown at severity level 3.}
    \label{fig:tiny_corruption_examples}
\end{figure}

\begin{figure}[h!]
    \centering
    \includegraphics[width=\linewidth]{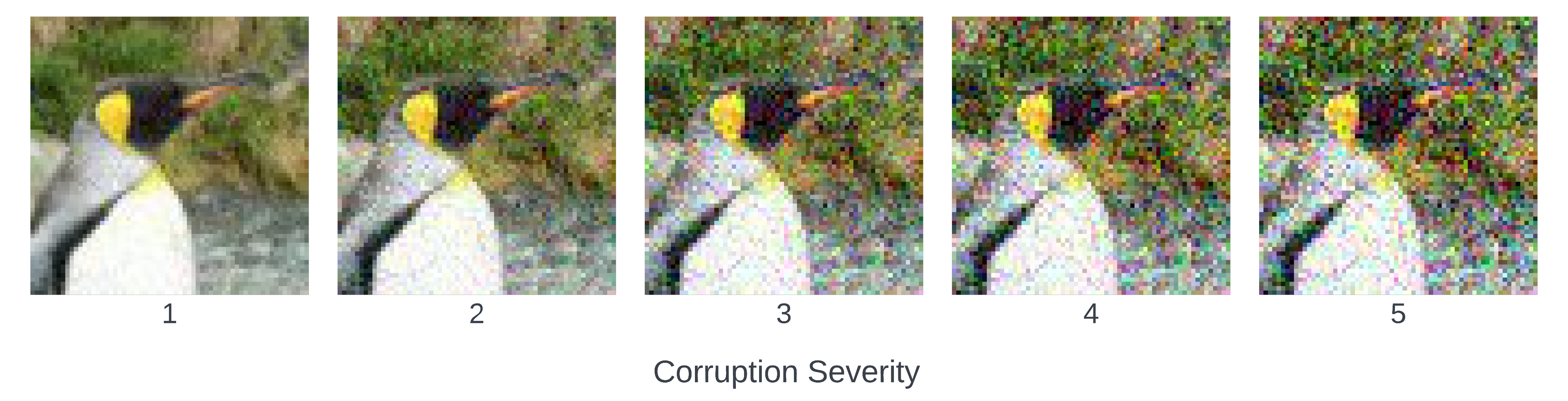}
    \caption{Gaussian noise corruption, shown at corruption severity levels 1-5.}
    \label{fig:tiny_corruption_severities}
\end{figure}

Tiny-ImageNet-C is among a collection of corrupted datasets (e.g., ImageNet-C, CIFAR-10-C, CIFAR-100-C) that feature a diverse set of corruptions to typical benchmark datasets.
Hendrycks and Dietterich~\citep{hendrycks_benchmarking_2019} suggest that given the diversity of corruptions featured in these datasets, performance on these datasets can be seen as a general indicator of model robustness.
The Tiny-ImageNet-C evaluation dataset consists of images from that Tiny-ImageNet validation dataset that have been corrupted according to $15$ types of image corruption, each of which is categorized as a `noise', `blur', `weather', or `digital' corruption.
The $15$ corruption types include: Gaussian noise, shot noise, impulse noise, defocus blur, frosted glass blur, motion blur, zoom blur, snow, frost, fog, brightness, contrast, elastic transformation, pixelation, and JPEG compression.
Each corruption is depicted in Fig.~\ref{fig:tiny_corruption_examples}.
Every image of this evaluation dataset is also corrupted at five levels of severity (the higher the corruption severity, the more the original image had been corrupted).
Corruption severities for Gaussian noise are exemplified in Fig.~\ref{fig:tiny_corruption_severities}.

\newpage

\newpage
\subsection{Corrupted Image Robustness}
A detailed breakdown of Tiny-ImageNet-C image classification accuracy for each single-component, neuro-constrained ResNet-50 and the composite models that achieved top V1 Overall score without adversarial training are provided in Tables \ref{tab:tiny_corruption_results_all}, \ref{tab:tiny_corruption_results_severity}, and \ref{tab:tiny_corruption_results_corruption_types}.

\begin{table}[h]
    \centering
    \setlength{\tabcolsep}{10pt} 
    \renewcommand{\arraystretch}{1.2}
    \resizebox{\linewidth}{!}{\begin{tabular}{lrrr}
        & Tiny-ImageNet Val. & Tiny-ImageNet-C & $\Delta$ \\
        \hline
        ResNet50 (Baseline) & $\mathbf{.742} \pm .003$ & $.278 \pm .004$ & $.463 \pm .006$ \\
        Center-surround antagonism & $.739 \pm .004$ & $.277 \pm .008$ & $.463 \pm .009$ \\
        Local Receptive Fields & $.741 \pm .002$ & $.275 \pm .004$ & $.467 \pm .004$ \\
        Tuned Normalization & $.740 \pm .001$ & $\mathbf{.283} \pm .006$ & $\mathbf{.457} \pm .006$ \\
        Cortical Magnification & $.683 \pm .001$ & $.222 \pm .009$ & $.461 \pm .009$ \\
        Composite Model A & $.694$ & $.231$ & $.463$ \\
        Composite Model B & $.691$ & $.232$ & $.459$ \\
    \end{tabular}}
    \caption{Classification accuracy of models on Tiny-ImageNet validation and Tiny-ImageNet-C (all corruption types and severities) datasets.  Composite Model A includes all 4 neuro-constrained architectural components (center-surround antagonism, local receptive fields, tuned normalization, and cortical magnification).  Composite Model B contained all architectural components, with the exception of center-surround antagonism. For baseline and single-component models, mean accuracies ($\pm$ one standard deviation) are reported, where each trial was associated with a distinct base model from the repeated trials of section \ref{section:individual-components}.}
    \label{tab:tiny_corruption_results_all}
\end{table}

\begin{table}[h]
    \centering
    \renewcommand{\arraystretch}{1.2}
    \resizebox{\linewidth}{!}{\begin{tabular}{lrrrrr}
        & \multicolumn{5}{c}{Corruption Severity} \\
        & 1 & 2 & 3 & 4 & 5 \\
        \hline
        ResNet50 (Baseline) & $.418 \pm .004$ & $.345 \pm .005$ & $.269 \pm .004$ & $.204 \pm .005$ & $.156 \pm .003$ \\
        Center-surround antagonism & $.414 \pm .010$ & $.343 \pm .009$ & $.267 \pm .009$ & $.203 \pm .006$ & $.156 \pm .004$ \\
        Local Receptive Fields & $.416 \pm .003$ & $.341 \pm .003$ & $.264 \pm .003$ & $.199 \pm .002$ & $.153 \pm .002$ \\
        Tuned Normalization & $\mathbf{.424} \pm .006$ & $\mathbf{.350} \pm .006$ & $\mathbf{.274} \pm .007$ & $\mathbf{.208} \pm .006$ & $\mathbf{.160} \pm .004$ \\
        Cortical Magnification & $.349 \pm .011$ & $.277 \pm .013$ & $.208 \pm .010$ & $.157 \pm .007$ & $.120 \pm .006$ \\
        Composite Model A & $.363$ & $.289$ & $.216$ & $.163$ & $.125$ \\
        Composite Model B & $.361$ & $.288$ & $.219$ & $.165$ & $.127$ \\
    \end{tabular}}
    \caption{Classification accuracy of models on Tiny-ImageNet-C at each level of corruption severity. Composite Model A includes all 4 neuro-constrained architectural components (center-surround antagonism, local receptive fields, tuned normalization, and cortical magnification).  Composite Model B contained all architectural components, with the exception of center-surround antagonism. For baseline and single-component models, mean accuracies ($\pm$ one standard deviation) are reported, where each trial was associated with a distinct base model from the repeated trials of section \ref{section:individual-components}.}
    \label{tab:tiny_corruption_results_severity}
\end{table}

\newpage
\begin{table}[h!]
    \centering
    \renewcommand{\arraystretch}{1.2}
    \resizebox{\linewidth}{!}{\begin{tabular}{lrrrr}
        \multicolumn{5}{c}{Noise Corruptions}\\
        & Gaussian Noise & Impulse Noise & Shot Noise & Avg. \\
        \hline
        ResNet50 (Baseline) & $\mathbf{.197} \pm .011$ & $.191 \pm .010$ & $\mathbf{.232} \pm .013$ & $\mathbf{.207} \pm .011$ \\
        Center-surround antagonism & $.195 \pm .010$ & $.186 \pm .009$ & $\mathbf{.232} \pm .012$ & $.204 \pm .010$ \\
        Local Receptive Fields & $.185 \pm .006$ & $.184 \pm 009$ & $.219 \pm .010$ & $.196 \pm .008$ \\
        Tuned Normalization & $.195 \pm .008$ & $\mathbf{.192} \pm .004$ & $.228 \pm .007$ & $.205 \pm .006$ \\
        Cortical Magnification & $.150 \pm .008 $ & $.157 \pm .007$ & $.180 \pm .011$ & $.162 \pm .008$ \\
        Composite Model A & $.151$ & $.156$ & $.184$ & $.164$ \\
        Composite Model B & $.144$ & $.149$ & $.177$ & $.157$ \\
        \\
    \end{tabular}}

    \resizebox{\linewidth}{!}{\begin{tabular}{lrrrrr}
        \multicolumn{6}{c}{Blur Corruptions}\\
        & Defocus Blur & Glass Blur & Motion Blur & Zoom Blur & Avg. \\
        \hline
        ResNet50 (Baseline) & $.224 \pm .003$ & $.182 \pm .001$ & $.272 \pm .003$ & $.241 \pm .004$ & $.230 \pm .002$ \\
        Center-surround antagonism & $.223 \pm .009$ & $.184 \pm .004$ & $.274 \pm .012 $ & $.243 \pm .011$ & $.231 \pm .009$ \\
        Local Receptive Fields & $.228 \pm .006$ & $.183 \pm .004$ & $.273 \pm .005$ & $.243 \pm .008$ & $.232 \pm .005$ \\
        Tuned Normalization & $\mathbf{.234} \pm .009$ & $\mathbf{.188} \pm .002$ & $\mathbf{.277} \pm .009$ & $\mathbf{.248} \pm .010$ & $\mathbf{.237} \pm .007$ \\
        Cortical Magnification & $.174 \pm .010$ & $.162 \pm .008$ & $.222 \pm .007$ & $.190 \pm .006$ & $.187 \pm .008$ \\
        Composite Model A & $.186$ & $.167$ & $.236$ & $.200$ & $.197$ \\
        Composite Model B & $.196$ & $.174$ & $.249$ & $.222$ & $.210$ \\
        \\
    \end{tabular}}

    \resizebox{\linewidth}{!}{\begin{tabular}{lrrrrr}
        \multicolumn{6}{c}{Weather Corruptions}\\
        & Brightness & Fog & Frost & Snow & Avg. \\
        \hline
        ResNet50 (Baseline) & $.401 \pm .005$ & $\mathbf{.282} \pm .003$ & $.360 \pm .006$ & $.310 \pm .004$ & $.338 \pm .004$ \\
        Center-surround antagonism & $.399 \pm .008$ & $.270 \pm .008$ & $.357 \pm .012$ & $.302 \pm .003$ & $.332 \pm .007$ \\
        Local Receptive Fields & $.398 \pm .008$ & $.275 \pm .005$ & $.351 \pm .006$ & $.298 \pm .004$ & $.331 \pm .003$ \\
        Tuned Normalization & $\mathbf{.410} \pm .008$ & $\mathbf{.282} \pm .011$ & $\mathbf{.361} \pm .006$ & $\mathbf{.311} \pm .010$ & $\mathbf{.341} \pm .008$ \\
        Cortical Magnification & $.327 \pm .011$ & $.211 \pm .013$ & $.283 \pm .014$ & $.248 \pm .010$ & $.267 \pm .011$ \\
        Composite Model A & $.338$ & $.220$ & $.286$ & $.258$ & $.275$ \\
        Composite Model B & $.327$ & $.225$ & $.284$ & $.255$ & $.273$ \\
        \\
    \end{tabular}}

    \resizebox{\linewidth}{!}{\begin{tabular}{lrrrrr}
        \multicolumn{6}{c}{Digital Corruptions}\\
        & Contrast & Elastic & JPEG & Pixelate & Avg. \\
        \hline
        ResNet50 (Baseline) & $.125 \pm .001$ & $.331 \pm .007$ & $.454 \pm .007$ & $.374 \pm .003$ & $.321 \pm .003$ \\
        Center-surround antagonism & $.122 \pm .002$ & $.331 \pm .014$ & $.455 \pm .007$ & $.374 \pm .004$ & $.321 \pm .006$ \\
        Local Receptive Fields & $.120 \pm .004$ & $.329 \pm .003$ & $.457 \pm .005$ & $.375 \pm .002$ & $.320 \pm .001$ \\
        Tuned Normalization & $\mathbf{.128} \pm .008$ & $\mathbf{.342} \pm .010$ & $\mathbf{.463} \pm .006$ & $\mathbf{.387} \pm .006$ & $\mathbf{.330} \pm .007$ \\
        Cortical Magnification & $.082 \pm .005$ & $.287 \pm .007$ & $.374 \pm .013$ & $.287 \pm .014$ & $.257 \pm .010$ \\
        Composite Model A & $.081$ & $.305$ & $.397$ & $.303$ & $.272$ \\
        Composite Model B & $.086$ & $.314$ & $.383$ & $.293$ & $.269$ \\
        \\
    \end{tabular}}
    \caption{Corrupted image classification accuracy by corruption type. 
 Composite Model A includes all 4 neuro-constrained architectural components (center-surround antagonism, local receptive fields, tuned normalization, and cortical magnification).  Composite Model B contained all architectural components, with the exception of center-surround antagonism. For baseline and single-component models, mean accuracies ($\pm$ one standard deviation) are reported, where each trial was associated with a distinct base model from the repeated trials of section \ref{section:individual-components}.}
    \label{tab:tiny_corruption_results_corruption_types}
\end{table}

%


\end{document}